\documentclass[preprint]{aastex}

\slugcomment{To appear in \emph{Astronomical Journal}, May, 2000}

\shortauthors{Matheson et al.}
\shorttitle{Helium Emission in the Type Ic SN 1999cq}

\begin{document}


\title{Helium Emission in the Type Ic SN 1999cq}

\author{Thomas Matheson, Alexei V. Filippenko, Ryan Chornock, 
Douglas C. Leonard, and Weidong Li}
\affil{Department of Astronomy, University of California,
    Berkeley, CA 94720-3411}

\email{matheson, alex, chornock, leonard, weidong@astron.berkeley.edu}


\begin{abstract}
We present the first unambiguous detection of helium emission lines in
spectra of Type Ic supernovae (SNe Ic).  The presence of \ion{He}{1}
lines, with full width at half maximum $\approx$ 2000 km s$^{-1}$, and
the distinct absence of any other intermediate-width emission (e.g.,
H$\alpha$), implies that the ejecta of SN Ic 1999cq are interacting
with dense circumstellar material composed of almost pure helium.
This strengthens the argument that the progenitors of SNe Ic are
core-collapse events in stars that have lost both their hydrogen and
helium envelopes, either through a dense wind or mass-transfer to a
companion.  In this way, SN 1999cq is similar to supernovae such as SN
1987K and SN 1993J that helped firmly establish a physical connection
between Type Ib and Type II supernovae.  The light curve of SN 1999cq
is very fast, with an extremely rapid rise followed by a quick
decline.  SN 1999cq is also found to exhibit a high level of emission
at blue wavelengths ($\lesssim$ 5500 \AA), likely resulting from
either an unusually large amount of iron and iron-group element
emission or uncharacteristically low reddening compared with other SNe
Ic.

\end{abstract}


\keywords{binaries: close---stars: evolution---stars:
mass-loss---supernovae: general---supernovae: individual (SN 1999cq)}

\section{Introduction}

The nature of supernovae (SNe) of Type Ib and Ic has been the subject
of much speculation; see Harkness \& Wheeler (1990) for an early
review, and Clocchiatti et al. (1997) for a more recent discussion.
Spectra of SNe Ib/c lack the hydrogen lines that distinguish SNe II,
yet they are also missing the deep absorption near 6150 \AA\ (thought
to be blueshifted \ion{Si}{2} $\lambda$6355) that characterizes SNe
Ia.  The defining characteristic of SNe Ib is the presence of strong
\ion{He}{1} lines near maximum light.  This is followed by a
nebular-phase spectrum dominated by emission lines of [\ion{O}{1}],
[\ion{Ca}{2}], and \ion{Ca}{2}.  SNe Ic are spectroscopically quite
similar to SNe Ib, but they do not show the strong \ion{He}{1} lines.
This results in the occasional designation of SNe Ic as
``helium-poor'' SNe Ib (Wheeler et al. 1987).

The most widely accepted model for the SN Ib/c explosion mechanism is
that it is related to the mechanism of SNe II---core collapse in
massive stars---but that SNe Ib/c have lost their hydrogen (and
helium, in the case of SNe Ic) envelopes through winds or mass
transfer to a companion (e.g., Woosley, Langer, \& Weaver 1993;
Yamaoka, Shigeyama, \& Nomoto 1993; Nomoto et al. 1994; Iwamoto et
al. 1994).  There is considerable circumstantial evidence that the
progenitors of SNe Ib and Ic are massive stars.  SNe Ib/c are
associated with Population I stars (Wheeler \& Levreault 1985; Uomoto \&
Kirshner 1985; Harkness et al. 1987; Huang 1987; Van Dyk 1992;
Bartunov et al. 1994, Van Dyk, Hamuy, \& Filippenko 1996), their hosts
are virtually always late-type galaxies (Porter \& Filippenko 1987),
and they exhibit radio emission (e.g., Weiler et al. 1998, and
references therein), thought to be the result of interaction with
circumstellar material (Chevalier 1982, 1984).  Contrasted with this
are the favored progenitors of SNe Ia, white dwarfs (e.g., Branch et
al. 1995; Livio 1999), although models for SNe Ib/c using white dwarfs
as progenitors have been proposed (e.g., Branch \& Nomoto 1986; Iben
et al. 1987).  The late-time spectra of SNe Ib/c are also very similar
to those of SNe II, with the obvious exception of the hydrogen lines.
Perhaps the best evidence connecting SNe Ib/c with SNe II is the
direct transformation of an individual supernova from one type to
another.  SN 1987K was initially spectroscopically a Type II, but the
late-time spectra resembled those of SNe Ib (Filippenko 1988); it thus
earned a new label, Type IIb (after Woosley et al. 1987).
Unfortunately, the actual transition was not observed.  Filippenko
(1992) and Jeffery et al. (1991) also suggested that the Type Ic SN
1987M showed evidence of hydrogen in its spectrum.

Like SN 1987K, SN 1993J was spectroscopically a SN II in the first few
weeks after explosion, but then began to exhibit the \ion{He}{1} lines
of a SN Ib, making it a SN IIb (Filippenko \& Matheson 1993;
Filippenko, Matheson, \& Ho 1993; Swartz et al. 1993a; Wheeler \&
Filippenko 1996).  The various theoretical models rapidly converged on
the same general concept---a massive star that had lost most, but not
all, of its hydrogen envelope (Woosley et al. 1994; Wheeler \&
Filippenko 1996, and references therein).  Gradually the [\ion{O}{1}],
[\ion{Ca}{2}], and \ion{Ca}{2} emission lines emerged, making the
spectrum of SN 1993J closely resemble that of an aging SN Ib, although the
hydrogen emission never fully disappeared.  Indeed, the hydrogen
emission became very prominent at late times, but this was almost
certainly the result of circumstellar interaction (Filippenko,
Matheson, \& Barth 1994).  Qiu et al. (1999) document thoroughly a
similar transition in SN 1996cb.

Although there has been no object showing as definitive a
transformation between SNe Ib and Ic as SN 1993J or SN 1996cb showed
between SNe II and Ib, there have been several SNe Ic with possible
signs of helium in their spectra.  These include the detection of
\ion{He}{1} $\lambda$10830 in SN 1990W (Wheeler et al. 1994; based on
this line, they propose SN 1990W was misclassified, and should be
Type Ib), SN 1994I (Filippenko et al. 1995), and SN 1994ai (Benetti,
quoted in Clocchiatti et al. 1997), in addition to the discovery of
other \ion{He}{1} lines at high velocity in a reanalysis of SN 1994I,
SN 1987M, and SN 1988L (Clocchiatti et al. 1996b) and in SN 1997X
(Munari et al. 1998).  (See, however, Baron et al. [1999] and Millard
et al. [1999] for cautionary arguments.)  By analogy to the models of
SN 1993J, one could postulate that the progenitors of SNe Ic have not
only had their hydrogen envelope removed, but most or all of their
helium layer as well (Harkness et al. 1987; Yamaoka, Shigeyama, \&
Nomoto 1993; Swartz et al. 1993b; Iwamoto et al. 1994; Nomoto et
al. 1994), so the weakness or absence of \ion{He}{1} lines is the
result of small abundance.  The SNe Ic that exhibited some helium
would then be the analogs of SN 1993J, but with the transition less
obvious.  An alternative explanation for the cause of the differences
between SNe Ib and Ic relies on the need for a large amount of
non-thermal excitation to form \ion{He}{1} lines (Harkness et
al. 1987; Lucy 1991).  The source of this excitation is
electrons accelerated by the $\gamma$-rays emitted by the decay of
radioactive $^{56}$Ni synthesized in the explosion.  In this scenario,
helium could be present in SNe Ic, but the $^{56}$Ni is not mixed into
the helium layers, and so cannot produce the helium lines
(e.g., Wheeler et al. 1987; Shigeyama et al. 1990; Hachisu et
al. 1991).  Whatever the cause of the helium features, the
characteristics of the previous detections of helium in SNe Ic
(absorption at high velocity) suggest that the helium is in the
ejecta, not the circumstellar material.

The spectrum of a truly transitional object between SNe Ib and Ic
could provide clues as to whether the difference between them is the
result of abundances, mixing, or both.  In this paper we present
optical to near-infrared (near-IR) spectra of SN 1999cq, a SN Ic with
unusual intermediate-width helium emission lines, along with an
unfiltered light curve.  We propose that the helium lines represent
the detection of material lost from the progenitor shortly prior to
explosion.  We also find an excess of blue emission in comparison with
other SNe Ic, with implications for iron and iron-group element
abundance, mixing, and the reddening of SN 1999cq.

\section{Observations}

SN 1999cq was discovered by the Lick Observatory Supernova Search
(LOSS) (Modjaz \& Li 1999) on 1999 June 25.4 UT, (JD 2,451,354.9; note
that all calendar dates used are UT) at an unfiltered magnitude of
$\sim$ 16.0.  It was present on previous images at similar magnitudes
( June 22.4, $\sim$ 15.9; June 19.4, $\sim$ 15.8).  An image taken on
1999 June 15.4 showed nothing at the position of the SN to a limit of
$\sim$ 19.0 mag.  The supernova was 1.$\arcsec$5 east and 4.$\arcsec$1
south of the nucleus of UGC 11268.  A spectrum obtained July 9.4 (JD
2,451,368.9, hereinafter ``the July spectrum'') indicated that SN 1999cq
was of Type Ib or Ic, and starting to enter the nebular phase,
although \ion{Na}{1}~D did show a P-Cygni profile (Filippenko 1999).

This spectrum was obtained with the Kast double spectrograph (Miller
\& Stone 1993) at the Cassegrain focus of the Shane 3-m reflector at
Lick Observatory with an exposure time of 1800 s.  Reticon $400
\times 1200$ pixel CCDs were used in both cameras.  The slit was
oriented at a position angle of 160$\arcdeg$ to include the galaxy
nucleus.  The optimal parallactic angle (Filippenko 1982) was
81$\arcdeg$, but the low airmass (1.1) and the large slit width
(3$\arcsec$) imply that the differential light losses were negligible.
Standard CCD processing and optimal spectral extraction were
accomplished with IRAF\footnote{IRAF is distributed by the National
Optical Astronomy Observatories, which are operated by the Association
of Universities for Research in Astronomy, Inc., under cooperative
agreement with the National Science Foundation.}.  We used our own
routines to flux calibrate the data, using the sdO comparison star
BD~+28${\arcdeg}$4211 (Stone 1977) in the range 3300--5400~\AA\ and
the sdG comparison star BD~+17${\arcdeg}$4708 (Oke \& Gunn 1983) in
the range 5300--9900~\AA.  Telluric absorption bands were removed
through division by the intrinsically featureless spectrum of
BD+17${\arcdeg}$4708 (Wade \& Horne 1988).  The dispersion of the
spectrum is $\sim$ 5 \AA\ per pixel.  We obtained a second spectrum on
1999 August 17.3 (JD 2,451,407.8, hereinafter ``the August spectrum'')
under almost identical conditions.  The only differences were a
narrower slit width (2$\arcsec$), a slightly higher airmass (1.2), and
the use of BD~+26${\arcdeg}$2606 (Oke \& Gunn 1983) as the comparison
star for the red wavelengths.

As SN 1999cq lies in a complex region only 4.$\arcsec$4 from the
nucleus of UGC 11268, background sky subtraction and the extraction of
the supernova were extremely difficult.  There is \ion{H}{2} region
contamination over a considerable portion of the long slit ($\sim$
48$\arcsec$ out of the $\sim$ 125$\arcsec$ spatial extent of our
exposures).  The least contamination of the object was achieved by
choosing sky regions for background subtraction very near the
supernova.  There is still some galaxy contamination, made more
complicated by unusually strong emission lines.  In fact, the regions
of the supernova spectrum near H$\alpha$+[\ion{N}{2}]
$\lambda\lambda$6548, 6583 and [\ion{S}{2}] $\lambda\lambda$6716, 6731
still show some unavoidable residual subtraction errors.  Incomplete
removal of the very strong \ion{Na}{1}~D night-sky emission line also
affects the spectrum.

The unusual nature of SN 1999cq was not immediately recognized, so it
was not followed photometrically through standard $UBVRI$ filters.
However, UGC 11268 is part of the LOSS sample of galaxies and is thus
imaged regularly (every three to four days) with the 0.76-m Katzman
Automatic Imaging Telescope at Lick Observatory (KAIT; Treffers et
al. 1997; Richmond, Treffers, \& Filippenko 1993).  The detector used
by KAIT is a SITe $512 \times 512$ pixel CCD with a field of view of
6.\arcmin8 $\times$ 6.\arcmin8.  The observations are taken without a
filter, so the light curve for SN 1999cq (Figure 1) is measured in
unfiltered magnitudes.  Photometric calibration of KAIT images
indicates that the unfiltered response is similar to a standard $R$
filter.  The images were reduced using a galaxy subtraction technique;
SN 1999cq is in a very complex region of the host galaxy, and
traditional aperture or point-spread-function (PSF) fitting photometry
would not perform well.  The template image of UGC 11268 was taken
1998 September 16 with a limiting magnitude of $\sim$ 20.  The
template is shifted spatially, scaled to the same intensity level and
PSF, and then subtracted from each individual observation.  Photometry
is then performed on the galaxy-subtracted images using PSF fitting in
DAOPHOT with more than 10 stars in the field used to construct the
PSF.  The magnitude of SN 1999cq was calculated by averaging its
magnitude relative to all the stars used to determine the PSF.  For
all of the observations of SN 1999cq the limiting unfiltered magnitude
was $\sim$ 19.

\section{Results}

Our spectra of SN 1999cq are shown in Figure 2.  A low value for the
Galactic component of reddening [$E(B-V) \approx 0.05$ mag] is found
using the Galactic maps of Schlegel, Finkbeiner, \& Davis (1998).
There is no obvious absorption near the location of \ion{Na}{1}~D at
the redshift of the host galaxy ($\sim$ 7900 km s$^{-1}$, determined
from the nuclear H$\alpha$ emission line; Marzke, Huchra, \& Geller
[1996] found 7890 $\pm$ 20 km s$^{-1}$ for UGC 11268).  To determine a
1$\sigma$ upper limit on the equivalent width (EW) of this line, we
used the relation EW(1$\sigma$) = $\Delta\lambda\ \times\ \Delta I$,
where $\Delta\lambda$ is the width of a resolution element (\AA) and
$\Delta I$ is the 1$\sigma$ root-mean-square (rms) fluctuation of the
spectrum around a normalized continuum level, based on the derivation
of Hobbs (1984).  The value we calculated for the \ion{Na}{1}~D line
in the July spectrum of SN 1999cq is EW $\lesssim$ 1.8 \AA.
The Barbon et al. (1990) relation would then give $E(B-V) \lesssim
0.45$ mag.  For reasons discussed below, however, we feel that the
intrinsic reddening is much lower than this upper limit.  We note that
the spectrum of the nucleus of UGC 11268 has a much higher
signal-to-noise ratio than that of the supernova, and it shows a
strong \ion{Na}{1}~D absorption line with an EW of $\sim$ 1.7 \AA, for
which the Barbon et al. (1990) relation gives $E(B-V) \approx 0.44$
mag.

Figure 3 provides a montage of SNe Ib and Ic at similar phases of
development for direct comparison with SN 1999cq.  While the July spectrum
of SN 1999cq in many ways resembles a typical SN Ic during the
transition to the nebular phase (e.g., emission lines of [\ion{O}{1}]
$\lambda\lambda$6300, 6364, [\ion{Ca}{2}] $\lambda\lambda$7291, 7324,
\ion{O}{1} $\lambda$7774, and the \ion{Ca}{2} near-IR triplet), there
are two striking differences.  The first is the presence in SN 1999cq
of several \ion{He}{1} lines in emission at an intermediate
velocity width compared with the other lines.  (Note that SN 1991ar
may show similar emission, but it is much less compelling.)
The second is the unusually large amount of emission at blue
wavelengths in SN 1999cq.

Most of the emission features in the July spectrum have a full width
at half-maximum (FWHM) of 7000-9000 km s$^{-1}$.  There are three
narrow, but resolved (FWHM $\approx$ 2000 km s$^{-1}$), lines that we
identify as \ion{He}{1} $\lambda$5876, $\lambda$6678, and
$\lambda$7065.  (A typical unresolved night-sky emission line has FWHM
$\approx$ 750 km s$^{-1}$).  The velocity width and flux are uncertain
for the \ion{He}{1} $\lambda$6678 line because of poor subtraction of
nearby [\ion{S}{2}] $\lambda\lambda$6716, 6731 emission (from
\ion{H}{2} regions) and also that its redshifted position of 6860 \AA\
is coincident with the telluric B-band.  In addition, \ion{He}{1}
$\lambda$5876 is superposed on the P-Cygni emission of \ion{Na}{1}~D.
These issues, along with the noisy nature of the spectrum, make flux
determinations problematic.  If we assign a flux value of 1.00 to
\ion{He}{1} $\lambda$5876, then \ion{He}{1} $\lambda$6678 is 0.65 and
\ion{He}{1} $\lambda$7065 is 1.61, with large uncertainties given
these caveats.

The August spectrum (Figure 2b) was extremely weak; the supernova had
faded considerably and was very difficult to detect.  (This was more
than 40 days past the last KAIT image where SN 1999cq was visible,
i.e. brighter than the limiting unfiltered magnitude of $\sim$ 19.)
Despite this, we still recover clear signs of the supernova.  The
\ion{Ca}{2} near-IR triplet is present, as is the strong amount of
emission at blue wavelengths that characterizes the July spectrum.
Most importantly, one of the helium lines (\ion{He}{1} $\lambda$7065)
is still clearly visible with approximately the same velocity width.

The photometric evolution of SN 1999cq is fairly unusual.  It is
difficult to make direct comparisons with other SNe as our values for
SN 1999cq are unfiltered magnitudes.  It rose rapidly, climbing above
the limiting value of 19 mag by 3.2 mag in 4 days.  The comparison
with the $R$-band rise of SN Ic 1994I (Richmond et al. 1996) in Figure
1 is particularly dramatic.  SN 1999cq then dropped quickly as well,
falling 0.2 mag in the 6 days after the first detection, and then 3.0
mag in the remaining 16 days that it was above the detection limit.
During a similar period (25 days) following maximum in the $R$ band,
SN 1994I fell 2.3 mag (Richmond et al. 1996), while SN 1993J declined
by 1.0 mag (Richmond et al. 1994).  If there are photometric
sub-classes for core-collapse SNe (Clocchiatti et al. 1996a, 1997),
then SN 1999cq clearly belongs to the ``fast'' class.  Models for SN
1994I have been able to reproduce a rapidly declining light curve with
a core-collapse event in stars whose envelopes have been lost to a
companion (Iwamoto et al. 1994; Woosley et al. 1995)

There have been no distance measurements for UGC 11268, although it is
a good candidate for IR Tully-Fisher distance determination (Haynes et
al. 1999).  If the recession velocity (7900 km s$^{-1}$) is chiefly
due the Hubble flow then the distance of UGC 11268 is 118 Mpc (using
$H_o = 67$ km s$^{-1}$ Mpc$^{-1}$), which is a distance modulus of
35.4 mag.  Although our photometric points are unfiltered magnitudes,
and the zero-point of our scale is not well determined, our brightest
observed magnitude of 15.8 and this distance modulus give us a rough
estimate of the absolute magnitude of SN 1999cq near maximum of
$-19.6$ mag.  While this value is quite uncertain, it is still
intriguing.  An absolute magnitude this bright would make SN 1999cq
one of the most luminous SNe Ic ever observed.  Clocchiatti et
al. (1999) report an absolute $V$ magnitude of $-19.2$ $\pm$ 0.02 or
$-20.2$ $\pm$ 0.02 for the Type Ic SN 1992ar.  (The two values are the
results of extrapolation of the light curve using two different
assumptions for its shape---the fainter value if it followed the
slow-type light curve, the brighter one if it were a fast-type
decliner.)  In addition, the Type Ic SN 1998bw (possibly associated
with gamma-ray burst 980425) had an absolute magnitude of $M_{V}$ =
$-19.35 \pm 0.05$ (Galama et al. 1998), with an extremely rapid rise
to maximum (Woosley, Eastman, \& Schmidt 1999).  As Clocchiatti et
al. (1999) discuss in detail, the existence of SNe Ic as bright as SNe
Ia complicates the identification of high-redshift SNe used to study
cosmological parameters (Riess et al. 1998; Perlmutter et al. 1999).
Without spectra of SN 1999cq near maximum light, it is not possible to
determine whether or not it could be mistaken for a SN Ia under the
difficult conditions presented by the observations of cosmologically
distant SNe.  Nevertheless, such luminous SNe Ic emphasize the need
for careful spectroscopic study of the high-redshift SNe.

\section{Discussion}

\subsection{Helium Lines}

The detection of the intermediate-width helium lines in the spectra of
SN 1999cq requires the presence of helium in the circumstellar
environment of the progenitor, but not in the ejecta of the exploding
star.  The width of the lines implies that the helium-emitting region
is interacting with the expanding ejecta.  The conspicuous absence of
Balmer emission suggests a very low abundance of hydrogen.  An obvious
solution is that SN 1999cq is interacting with material from the
progenitor lost through a wind or mass transfer to a companion.

It is also possible that the source of the helium lines is material
not originally from the SN progenitor, but rather a companion
sufficiently nearby to be affected by the explosion.  As mentioned
before, mass transfer facilitated by a companion is a common model for
SNe Ib/c progenitors, and thus the presence of such a companion must
be considered.  We note that Chugai (1986) and Marietta, Burrows, \&
Fryxell (1999) predicted a similar velocity width ($\lesssim$ 1000 km
s$^{-1}$) for the lines of hydrogen from the interaction of a SN Ia
with its companion, although those calculations were for material
entrained in the ejecta and the resulting emission would not be
visible until several hundred days after the explosion.  For SN
1999cq, however, such a companion star would probably be hydrogen-rich
if it had accreted hydrogen from the progenitor; thus, it is more
likely that the helium lines are evidence of helium-rich material that
was lost from the progenitor itself prior to core collapse.

The helium lines exhibit unusual line intensity ratios.  In a typical
hydrogen-rich atmosphere with a temperature of $(1-2) \times10^4$ K
and electron density $n_e \approx 10^2-10^6$ cm$^{-3}$ one would expect the
ratio of \ion{He}{1} $\lambda$7065 to \ion{He}{1} $\lambda$5876 to be
$\sim$ $0.1-0.2$ (Osterbrock 1989).  The ratio observed here (1.61)
implies a very different state, most likely related to the
non-local-thermodynamic-equilibrium excitation required for the helium
emission (Lucy 1991).  The work of Almog \& Netzer (1989), however,
studying the emission-line spectrum of \ion{He}{1} over a wide range
of physical conditions, indicates that the large 7065/5876 ratio could
be the result of a very high density.  We cannot directly apply the
numbers calculated by Almog \& Netzer, as their models were for helium
in a hydrogen atmosphere.  Nonetheless, their results indicate that
for a small range of densities ($n_e \approx 10^{10}$ cm$^{-3}$, at $T
= 10^4$ K), the strength of the 7065 line becomes greater than that of
5876.  Such a large ratio of 7065/5876 has been observed in other SNe
that exhibited circumstellar interaction through helium and hydrogen
emission lines, including intermediate-width lines (e.g., SN 1996L;
Benetti et al. 1999).  Assuming that the trend of increasing 7065/5876
with increasing density applies to an effectively pure helium
atmosphere, then the intermediate-width helium lines in SN 1999cq
originate in an extremely dense environment.

Hydrogen and helium emission lines of similar width were also present
in the spectra of SN 1988Z (Filippenko 1991; Stathakis \& Sadler 1991;
Turatto et al. 1993).  Chugai \& Danziger (1994) (hereinafter CD94)
interpreted this intermediate-width component of the emission as
evidence for slow radiative shocks propagating either in dense clumps
of wind material or in a dense equatorial belt, with both embedded in
a uniform wind.  Unfortunately, without multiple epochs and well
defined broad lines, it is not possible to apply the details of CD94's
models.  Moreover, SN 1999cq has the added complication of an
apparently pure helium wind, where the issues of excitation and
radiative transfer are not as well known; such a calculation is beyond
the scope of this paper.  Therefore, we cannot differentiate between
the two scenarios of CD94.  Nonetheless, the qualitative arguments
still apply.  Intermediate-width components in the spectra of SNe are
likely to be the result of ejecta interacting with dense material lost
from the progenitor prior to explosion.  The lack of narrow
(unresolved) and broad (FWHM $\gtrsim$ 5000 km s$^{-1}$) \ion{He}{1}
lines in our spectra of SN 1999cq implies that the uniform component
of both of CD94's models is either extremely tenuous and/or that
helium excitation is more difficult in such rarefied environments.

As noted above, several models for the progenitors of SNe Ib and Ic
postulate that they are massive stars that lose their envelopes prior
to core collapse.  Both SN 1987K and SN 1993J provided direct evidence
for this mechanism by exhibiting hydrogen emission from the low-mass
layer of the progenitor's original envelope that remained.  Because
the late-time spectra of SNe Ib and Ic are similar (see, e.g.,
Filippenko 1997), the analogous transformation of a SN Ib to a SN Ic
would be much more subtle.  In fact, SNe that have broad He absorption
lines that fade during the transition to the nebular stage are, by
definition, still called SNe Ib.  SN 1999cq thus provides evidence for
a link in another way.  The \ion{He}{1} lines we observe indicate an
almost pure helium mass loss---exactly what the above models have
assumed to produce the stripped-envelope progenitors of SNe Ic.

One question about SN 1999cq remains: Is it legitimately a SN Ic?  The
late-time spectra of SNe Ib and Ic are similar overall, but the
velocity-widths of the lines are different.  Schlegel \& Kirshner
(1989) found the widths of [\ion{O}{1}] $\lambda\lambda$6300, 6364 and
[\ion{Ca}{2}] $\lambda\lambda$7291, 7324 in late-time spectra of SNe
Ib to be FWHM = 4500 $\pm$ 600 km s$^{-1}$.  In contrast, Filippenko
et al. (1995) reported FWHM values for [\ion{Ca}{2}] of 9200 km
s$^{-1}$ for SN Ic 1994I and 6200 km s$^{-1}$ for SN Ic 1987M, for
spectra at similar phases ($\sim$ 4.6 months after maximum).  For
[\ion{O}{1}], the FWHM was 7700 km s$^{-1}$ for SN 1994I and 7500 km
s$^{-1}$ for SN 1987M.  In SN 1999cq, the FWHM of the [\ion{O}{1}]
line is 7800 km s$^{-1}$, while the [\ion{Ca}{2}] line is 8000 km
s$^{-1}$ (given the difficulty of determining a continuum level, the
uncertainty of these values is $\sim$ 15\%).  The spectrum of SN
1999cq was obtained only a few weeks past maximum, but its rapid
evolution (c.f. Figure 1) implies that the relative phase is at least
comparable.  Filippenko et al. (1995) interpreted the larger
velocity-width of the lines in SN 1994I as indicative of either a
higher explosion energy and/or a smaller ejected mass in comparison
with SN 1987M.  Following this interpretation, SN 1999cq would have an
even larger explosion energy and/or smaller ejected mass.  This is
consistent both with the evidence for mass loss, as well as the large
luminosity of the event, and it makes the classification of SN 1999cq
as a Type Ic event more secure.

Another difference between SNe Ib and SNe Ic is the shape of their
light curves.  The light curves of SNe Ib decline more slowly,
typically dropping only $\sim$ 1.5 mag. in the 25 days after maximum
($B$ magnitude; Schlegel \& Kirshner 1989).  As noted before, in that
same time SN 1994I dropped 2.3 mag. and SN 1999cq fell by 3.2 mag.
This implies an even more rapid transition to the nebular phase than
that of SN 1994I ($\lesssim$ 2 months; Filippenko et al. 1995).  The
first spectrum (July) was obtained only 24 days following the most
recent image without the supernova (the temporal position relative to
the light curve is marked in Figure 1), and it clearly shows the onset
of nebular features.  The rapid rise and fall of SN 1999cq also imply
that it was of Type Ic.

There is no evidence in the spectra of SN 1999cq for the
distinguishing characteristic of SNe Ib---broad helium lines,
generally interpreted as P-Cygni profiles.  We believe that the
P-Cygni profile of \ion{Na}{1}~D is still evident in Figure 2.  The
relatively large amount of blue emission in SN 1999cq makes the
P-Cygni absorption trough more difficult to discern, but it is there.
The blue minimum of the line indicates an expansion velocity of 8300
km s$^{-1}$, well in agreement with the widths of the broad lines.  If
SN 1999cq did have broad He lines initially, they must have faded very
quickly, while the \ion{Na}{1}~D line remained.  Nonetheless, even if
SN 1999cq were originally a SN Ib, the intermediate-width \ion{He}{1}
lines still imply pure helium mass loss.

\subsection{Blue Emission}

As Figure 3 shows, SN 1999cq also exhibits an anomalous ``blue bump''
of emission in contrast to other SNe Ic.  This appears in both of our
spectra (July and August).  The spectra of the host galaxy
nucleus and nearby \ion{H}{2} regions extracted from the same
long-slit exposures show no evidence for unusual blue emission.  Given
the extreme overlapping of lines, it is difficult to identify
individual features other than perhaps \ion{Mg}{1}] $\lambda$4571 and
some lines of \ion{Fe}{2}.

One possible source for the blue emission is a mixture of overlapping
iron lines.  Clocchiatti et al. (1997) identified the \ion{Fe}{2}
lines at $\lambda$4924, $\lambda$5018, and $\lambda$5169 in the Type
Ic SN 1983V.  They also found \ion{Fe}{2} blends of multiplets 27 and
28 ($\sim$ $4100-4400$ \AA), multiplets 37 and 38 ($\sim$ $4500-4650$
\AA), and multiplet 74 (near 6200 \AA).  Other possibilities include
multiplets 3, 14, and 29 ($\sim$ $3800-4000$ \AA), multiplets 43 and
50 ($\sim$ $4700-4750$ \AA), multiplet 42 ($\sim$ $4900-5200$ \AA),
and multiplets 48 and 49 ($\sim$ $5200-5400$ \AA) (e.g., Phillips
1976).  There is clearly no dearth of potential iron lines in the
region of our spectrum that shows this blue emission.  In addition,
there are several lines from other stable iron-group elements such as
Co and Ni (see, e.g., Axelrod 1980).

The presence of such strong iron (and/or iron-group) lines in SN
1999cq compared to other SNe Ic might imply that the mixing of
$^{56}$Ni was more extensive in SN 1999cq.  This would, however,
contradict the standard predictions of the mixing differences between
SNe Ib and Ic (Baron 1992), but only if helium were still present in
the atmosphere of the progenitor.  If the progenitor of SN 1999cq had
lost most or all of its helium, leaving a C-O core, then considerable
mixing of $^{56}$Ni to the outer layers is possible, resulting in
excess iron emission.  This could still be consistent with models for
mixing in SNe Ic as the $^{56}$Ni is not reaching beyond the C-O core
into a helium layer.  In addition, the bright absolute magnitude of SN
1999cq may imply the production of more $^{56}$Ni than is produced in
a typical SN Ic, thereby providing the excess material to mix into the
outer parts of the ejecta.

Given the relatively large cross-sections for the iron-group elements,
however, the strong emission lines might not require overly large
abundances. A fully quantitative spectral synthesis model would be
necessary to determine the actual abundances.  A further concern for
the effect of mixing on the character of the spectrum is that
macroscopic mixing of $^{56}$Ni could occur without necessarily
exciting helium, and thus produce the apparent excess of iron-group
element emission without creating broad helium lines, as is observed
in SN 1999cq.  Therefore, SN 1999cq may not provide a discriminant
between the mixing models for SNe Ib and Ic.  While there are
variations in abundances (mainly light elements) and light-curve
shapes (which have implications for synthesized $^{56}$Ni) for SNe Ib
and SNe Ic; e.g., Clocchiatti et al. 1997, and references therein), to
our knowledge no other SN Ib or Ic has shown such extreme iron or
iron-group emission.  Spectra of other luminous SNe Ic (SN 1992ar,
Clocchiatti et al. 1999; SN 1998bw, Branch 1999) do not show unusual
emission at blue wavelengths.

Another explanation for the ``blue bump'' is related to the
surrounding environment of the supernova.  One way to increase the
blue emission of an object is to scatter the spectrum off of
interstellar dust, as in a light echo.  This phenomenon has a long
history of discussion in association with SNe (e.g., Zwicky 1940; van
den Bergh 1975; Chevalier 1986; Schaefer 1987).  It has only been
definitively observed in two supernovae: SN 1987A (Suntzeff et
al. 1988; Crotts 1988; Gouiffes et al. 1988) and SN 1991T (Schmidt et
al. 1994; Sparks et al. 1999).  With spectra from earlier epochs, one
can match late-time echoes.  Suntzeff et al. (1988) and Schmidt et
al. (1994) each characterized the effects of scattering as a power law
($F_\lambda \propto \lambda^{-\alpha}$), with $\alpha$ values of 4.9
$\pm$ 0.8 and 2, respectively.  (Crotts [1988] found $\alpha = 3.5 \pm
0.5$ from broad-band colors.)  Not having other spectra of SN 1999cq,
we instead compared it with spectra of the SNe Ic 1987M and 1994I that
had been corrected by the scattering law.  We find that
$\alpha\thinspace\approx 1.5$ gives a good fit.  This variation of the
exponent in the scattering power law is presumably related to the
nature of the scattering dust.  The scattering efficiency for
particles of a size scale comparable to optical wavelengths decreases
with decreasing size (although it will also decrease for particles
much larger than optical wavelengths, eventually producing grey
extinction), but the shape and composition of the particles has a
significant enough effect on scattering to preclude much speculation
regarding the nature of the dust (e.g., Yanamandra-Fisher \& Hanner
1999).

As reddening is, in practice, the inverse process to scattering, a
more likely solution related to environmental effects is a lack of
extinction for SN 1999cq.  Both SN 1994I and SN 1987M were
considerably reddened.  For SN 1994I, Ho \& Filippenko (1995) found
$E(B-V) = 1.0^{+1.0}_{-0.5}$ mag (assuming $R$ = $A_V/E(B-V)$ = 3.0)
from high-resolution studies of the \ion{Na}{1}~D lines, but they
considered $E(B-V) \lesssim 0.47$ mag a more likely limit; this is
also the value Iwamoto et al. (1994) found from light-curve studies.
Filippenko, Porter, \& Sargent (1990) estimated the reddening of SN
1987M to be $E(B-V) \approx 0.44$ mag.  The spectrum of SN 1995F in
Figure 3 also shows a strong \ion{Na}{1}~D absorption feature (EW
$\approx 1.0$ \AA) that implies an $E(B-V) \approx 0.25$ mag from the Barbon
et al. (1990) relation.  Dereddening these SNe using the extinction
correction of Cardelli, Clayton, \& Mathis (1989), with the O'Donnell
(1994) modifications at blue wavelengths, results in spectra that
effectively match the ones produced by correcting with the previously
described scattering law (Figure 4).

As mentioned above, the intrinsic reddening of SN 1999cq is highly
uncertain.  The 1$\sigma$ upper limit for $E(B-V)$ of 0.45 mag is
reasonable when compared with other SNe Ic.  A more quantitative
estimate, using techniques to be described elsewhere (Leonard et al.,
in preparation), is that the EW is consistent with zero, but with a
1$\sigma$ uncertainty of 0.97 \AA.  Thus the reddening of $E(B-V)
\lesssim 0.45$ mag is probably an overestimate, with $E(B-V) \lesssim
0.25$ being more likely.  Given how well the dereddened spectra of
other SNe Ic match SN 1999cq, it is plausible that it is not
significantly reddened.  We believe that what appears to be an
anomalous level of blue emission might actually represent the true
spectrum of an unreddened SN Ic, perhaps with unusually strong
\ion{Fe}{2} emission since SN 1999cq was such a luminous event.
Without a reliable estimate of the extinction, though, this conclusion
is highly uncertain.

\section{Conclusions}

Spectra of SN 1999cq reveal significant emission at blue wavelengths
in excess of what is observed in other SNe Ic.  Any interpretation of
this emission is strongly dependent on the intrinsic reddening of SN
1999cq.  If we assume SN 1999cq suffers little reddening, then its
spectra are indicators of the relative significance of blue emission,
presumably iron and iron-group lines, in SNe Ic.  The blue emission
would contribute almost equally with the standard red nebular lines
([\ion{O}{1}],[\ion{Ca}{2}], and \ion{Ca}{2}), in contrast to typical
SNe Ic.  If SN 1999cq is as reddened as SNe Ic usually are, then the
alternative solution for the excess blue emission of unusually strong
iron and iron-group element emission would be just as interesting,
with potential implications for the mixing of $^{56}$Ni into the outer
parts of the progenitor and/or the total amount of $^{56}$Ni produced
by the explosion.  The detailed nature of the mixing of nickel must be
understood to see if mixing models can actually produce the difference
between SNe Ib and SNe Ic.  If the helium is not present, mixing could
still occur, but then the abundance of helium is the critical factor
in determining whether or not the core collapse of a stripped star is
observed as a SN Ib or a SN Ic.

We have also discovered \ion{He}{1} emission lines with an
intermediate velocity width in SN 1999cq, a SN Ic.  Since such
intermediate-width lines likely indicate the interaction of the SN
ejecta with dense material lost from the progenitor star, we believe
this provides evidence of an almost pure helium wind or mass transfer.
If this is the case, then the mechanism that differentiates SNe Ib and
Ic is similar to the mass loss or transfer that transforms a Type II
core-collapse SN into a Type Ib/c.  The \ion{He}{1} lines in SN 1999cq
have the signature of dense mass loss of a helium envelope, which
would not be expected if differing degrees of mixing of $^{56}$Ni into
the ejecta were the sole criterion for the difference between SNe Ib
and Ic.  SN 1999cq does not show a direct transformation such as that
of SN 1987K, SN 1993J, or SN 1996cb, but, in the \ion{He}{1} lines, we
may be seeing a reflection of the last step in the process that
creates the progenitors of SNe Ic.

\acknowledgments This research was supported by NSF grant AST-9417213.
We are grateful to the staff of Lick Observatory (especially K. Baker,
W. Earthman, and A. Tullis) for their assistance with the
observations.  We also thank A. L. Coil, J. R. Graham, and
J. C. Shields for helpful discussions.  We are grateful to Sun
Microsystems Inc. (Academic Equipment Grant Program), Hewlett-Packard
Inc., the National Science Foundation, and the Sylvia and Jim Katzman
Foundation for donations that made KAIT possible.


\clearpage



\figcaption{Light curve of SN 1999cq.  Filled circles represent
unfiltered magnitudes (approximately $R$ band) of the supernova; open
circles represent the limiting magnitude for images in which the
supernova was not visible.  The error bars are estimates, not formal
uncertainties for the magnitudes.  Errors for the first four
magnitudes of SN 1999cq are smaller than the size of the points.  The
crosses are the $R$-band light curve of SN 1994I (Richmond et al. 1996,
Leuschner 0.5-m data only) scaled and shifted to match approximately
the maximum value for SN 1999cq.  The date of the July spectrum is
also indicated.\label{fig1}}

\figcaption{Spectra of SN 1999cq from (a) 1999 July 9.4 and (b) August
17.3 with some line identifications.  A recession velocity of 8200
km s$^{-1}$, as determined from the centroid of the intermediate-width
\protect\ion{He}{1} lines, has been removed. (The host galaxy nucleus has a
recession velocity of 7900 km s$^{-1}$.)  Significantly poor
subtraction of H$\alpha$ from nearby \protect\ion{H}{2} regions has been
deleted from spectrum (b).  Note the continuing presence of
\protect\ion{He}{1} $\lambda$7065, the \protect\ion{Ca}{2} near-IR
triplet, and the excess blue emission in spectrum (b).\label{fig2}}

\figcaption{Spectra of various SNe (SN Ib 1991ar; SNe Ic 1987M, 1994I,
and 1995F) at a similar phase of development as the July spectrum of
SN 1999cq.  (The SN 1991ar spectrum is 30 days past discovery
[McNaught \& Russell 1991] with no information relative to maximum;
the SN 1987M spectrum is 96 days past $B$ maximum [Filippenko et
al. 1990]; the SN 1994I spectrum is 56 days past $B$ maximum
[Filippenko et al. 1995]; and the SN 1995F spectrum is 74 days past
the earliest image [Vagnozzi, Piermarini, \& Russo 1995] with no
information relative to maximum).  All have been deredshifted by the
appropriate velocity and narrow emission lines have been removed,
where applicable.  The spectra show substantial similarity, the only
major variation among them being the relative ratios of the features
in the range $\sim$ $6100-6600$ \AA.  The broad lines in this region
include [\protect\ion{O}{1}] $\lambda$6300, and possibly
\protect\ion{Si}{2} $\lambda$6355 and \protect\ion{C}{2} $\lambda$6580
(see, e.g., Baron et al. 1999; Millard et al. 1999).  The dramatic
differences that SN 1999cq exhibits are the intermediate-width
\protect\ion{He}{1} lines and the considerable emission blueward of
5500 \AA.  Note, however, that SN 1991ar may also have
intermediate-width \protect\ion{He}{1} lines which had not been
noticed previously.  In addition, the \protect\ion{O}{1} $\lambda$7774
line in SN 1999cq appears to be redshifted when compared with the
other SNe.  This is not evident in any other lines.  We have no
explanation for this anomaly, but we believe it does not affect our
interpretation of the rest of the spectrum.
\label{fig3}}

\figcaption{Spectrum of SN 1999cq (smoothed with a 3 pixel-width
boxcar to emphasize the overall shape) compared with spectra of SN
1994I that have been (a) corrected for scattering, (b) corrected
for reddening, and (c) uncorrected.  The scattering correction was
$F_\lambda \propto \lambda^{-1.5}$.  Dereddening by $E(B-V) = 0.47$
mag was accomplished with the extinction corrections of Cardelli et
al. (1989), including the O'Donnell (1994) modifications.  Note that
both the (a) and (b) versions of SN 1994I have distinctly different
overall shapes when compared with (c).  The increased level of blue
emission evident in the corrected spectra of SN 1994I is a fairly good
match to the amount of anomalous blue emission of SN 1999cq.  The
slope of the underlying continuum of SN 1999cq is clearly different
from the observed spectrum (c) of SN 1994I.\label{fig4}}

\clearpage


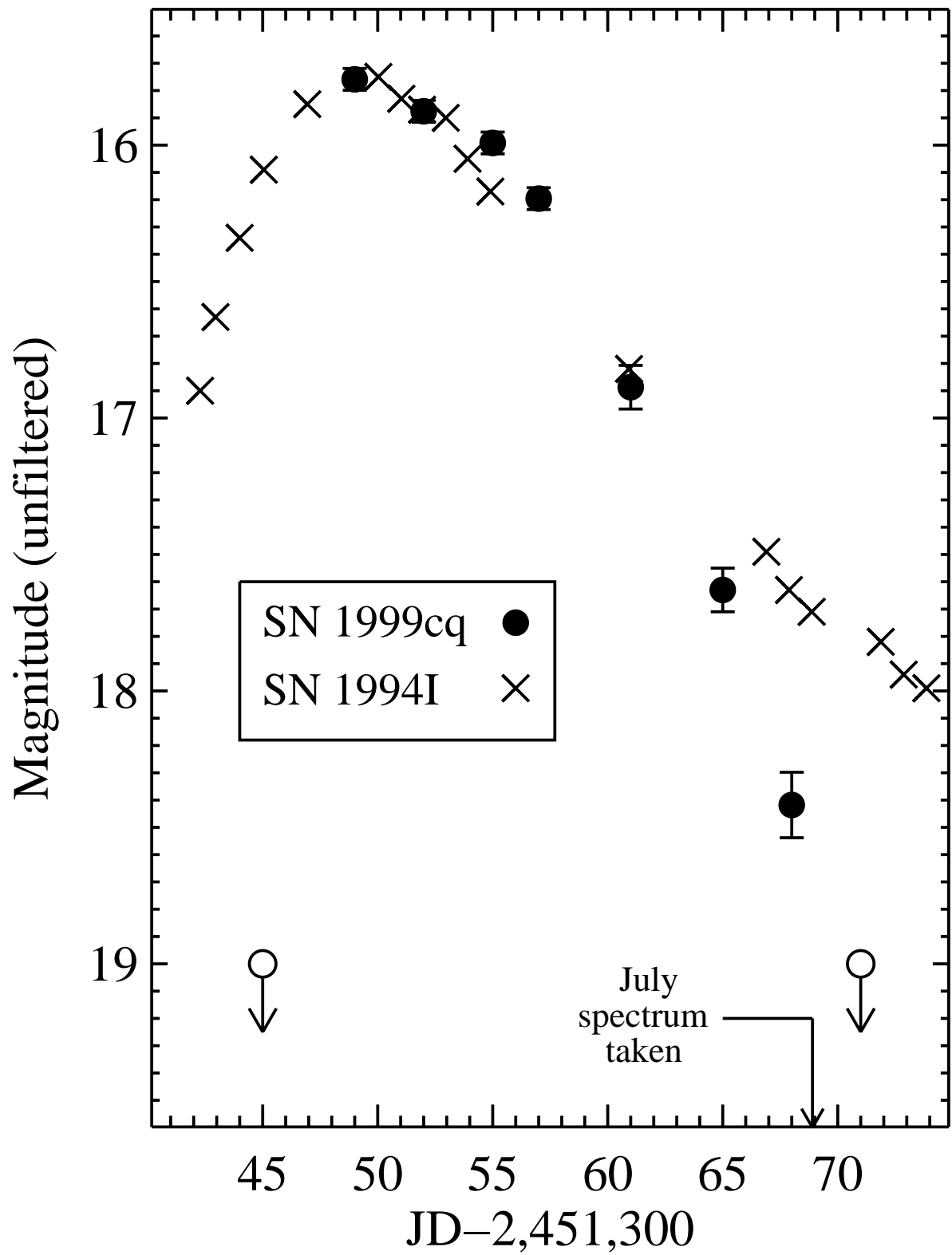
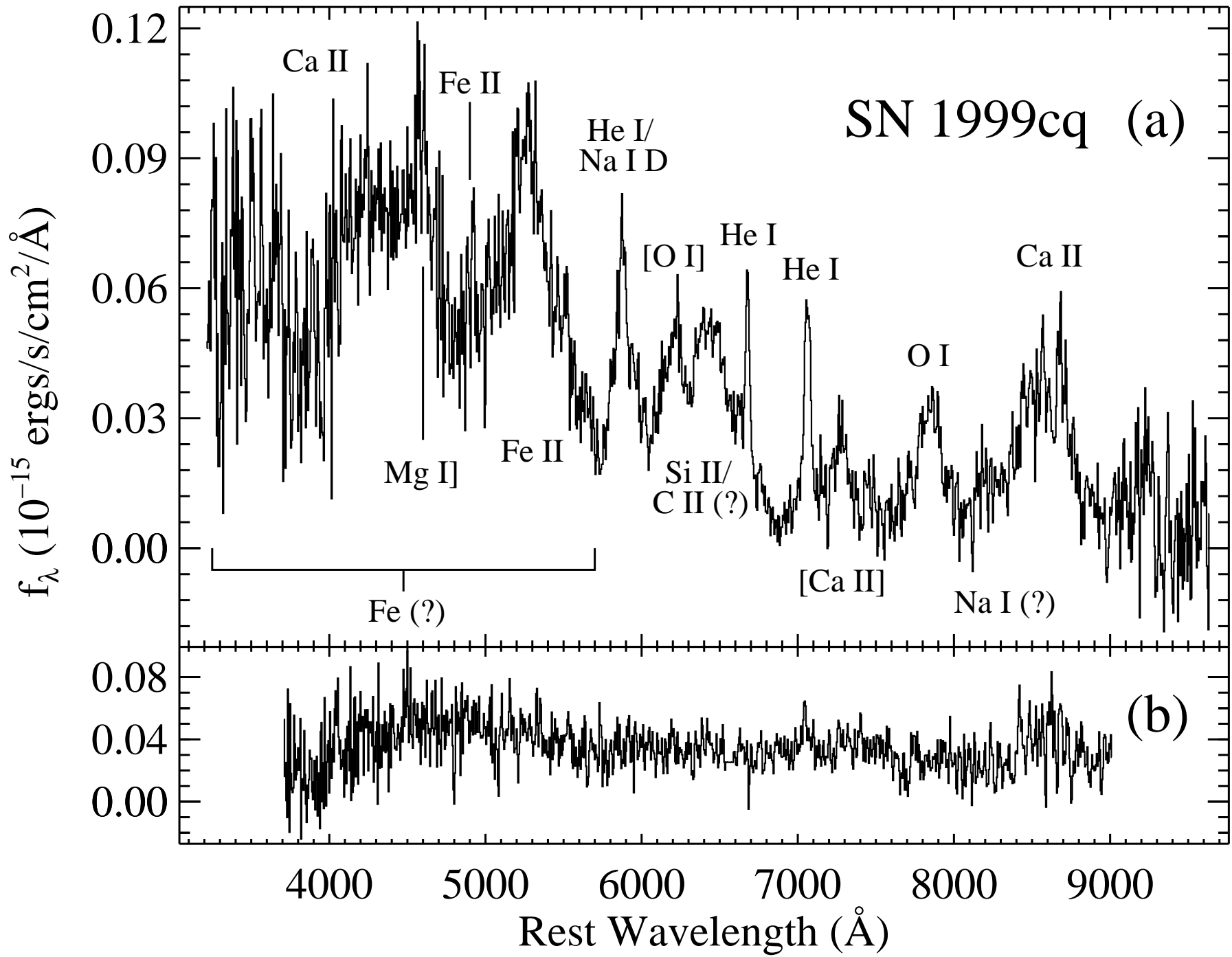
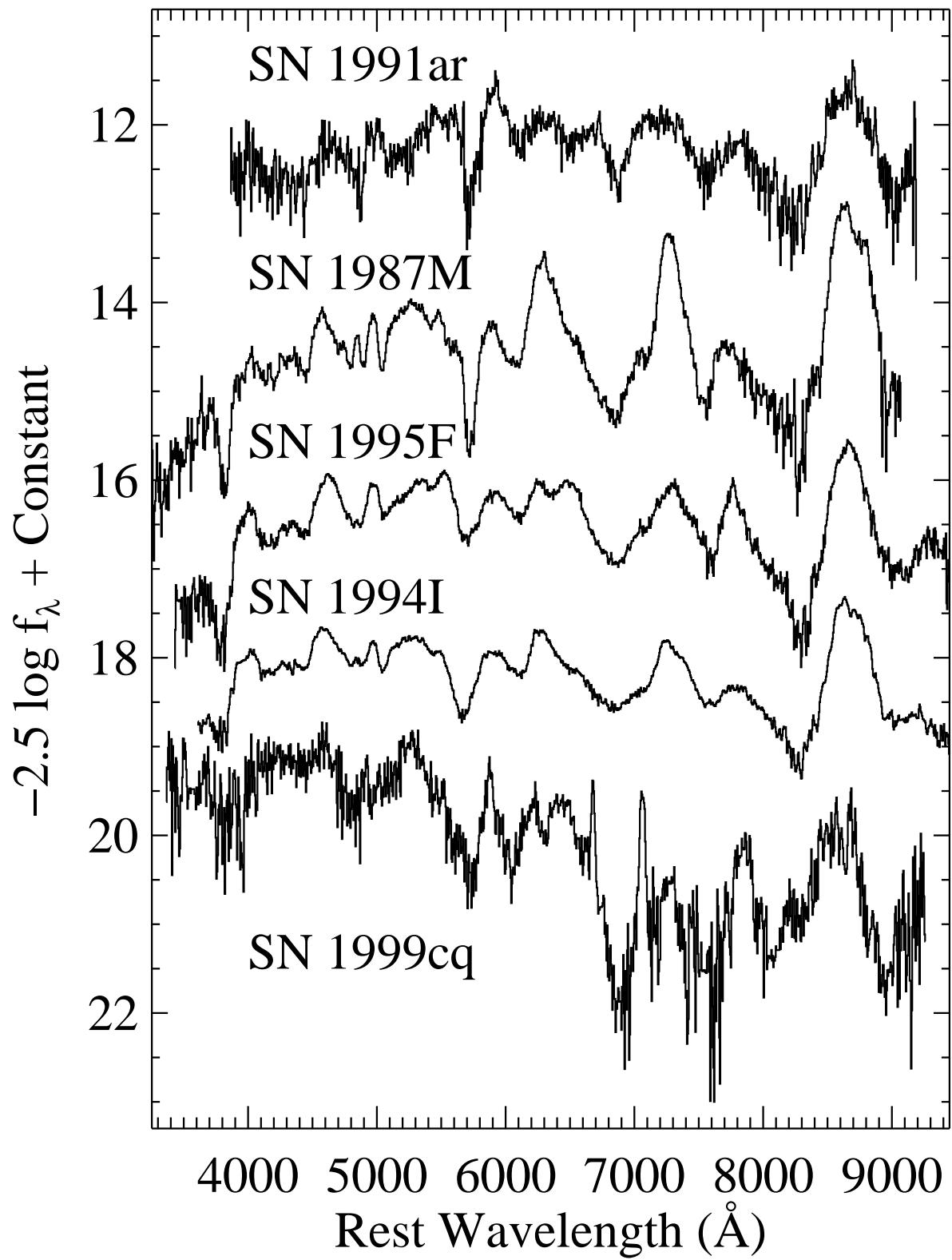
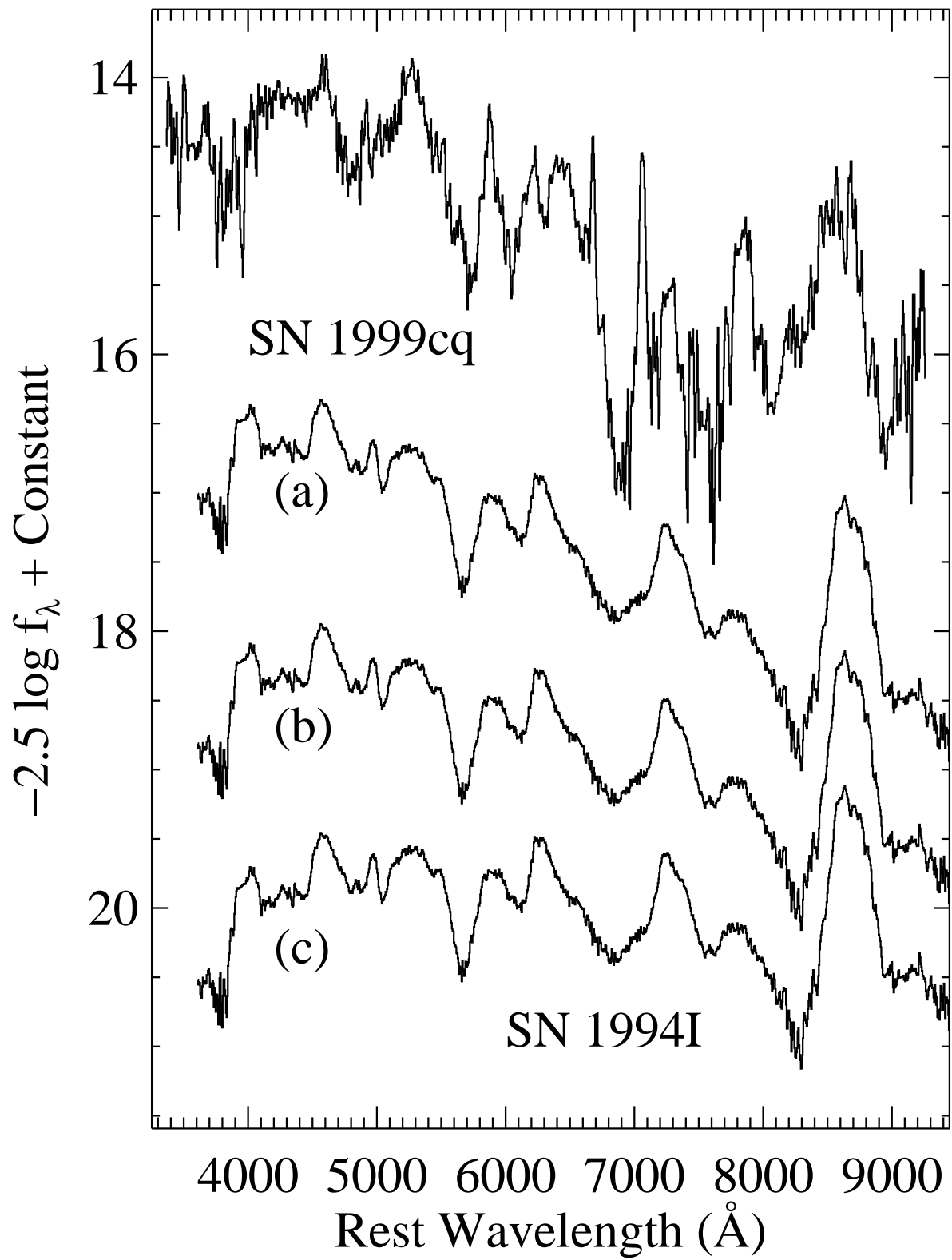
\end{document}